\documentclass[aps,prl,twocolumn,groupedaddress]{revtex4}

\usepackage{graphicx}
\usepackage{dcolumn}
\usepackage{bm}
\usepackage[english]{babel}
\usepackage[applemac]{inputenc}
\usepackage{amsmath}

\newcommand\BA{\begin{align}}
\newcommand\EA{\end{align}}
\newcommand\BE{\begin{equation}}
\newcommand\EE{\end{equation}}
\newcommand\BF{\begin{figure}}
\newcommand\EF[2]{\caption{#1}\label{#2}\end{figure}}
\def\II{1\!\mathrm{l}}
\def\cZ{\mathcal Z}

\def\ket#1{| #1 \rangle}

\begin{document}

\title{Fast Decoders for Topological Quantum Codes}

\author{Guillaume Duclos-Cianci}
\author{David Poulin}

\affiliation{D\'epartement de Physique, Universit\'e de Sherbrooke, Qu\'ebec, Canada}

\date{\today}

\begin{abstract}
We present a family of algorithms, combining real-space renormalization methods and belief propagation, to estimate the free energy of a topologically ordered system in the presence of defects. Such an algorithm is needed to preserve the quantum information stored in the ground space of a topologically ordered system and to decode topological error-correcting codes. For a system of linear size $\ell$, our algorithm runs in time $\log\ell$ compared to $\ell^6$ needed for the minimum-weight perfect matching algorithm previously used in this context and achieves a higher depolarizing error threshold.
\end{abstract}

\maketitle

Topologically ordered phases of matter can be used to store and process quantum information in an inherently robust way \cite{Kit97c,Kit03a,FKLW02a,DKLP02a,B06a}. The ground state degeneracy depends on the topology of the system. Quantum information stored in this ground state manifold is protected from local perturbations because virtual transitions require an order in perturbation proportional to the linear size of the system. 

At finite temperature, thermal excitations create pairs (or larger sets)  of particles  of finite mass. These particles are not confined and can freely diffuse on the surface. Bringing all the particles to fuse in pairs returns the system into a ground state. This ground state will correspond to the original one only if the world-line of the particles has a trivial homology: otherwise the particles have generated a topological transformation that corrupts the information stored in the ground space. Since particles appear in a finite density ($e^{-\frac{\rm{mass}}{k_BT}}$) at any non-zero temperature, this information corruption happens on a time-scale independent of the system size \cite{NO08a}.

To store the information for longer times, it becomes necessary to keep track of the thermal particles and have accurate knowledge of their world-line homology. This is possible if the locations of the particles are measured repeatedly on a time-scale shorter or comparable to their diffusion rate. It is also necessary to process the information gathered from these measurements rapidly, i.e., to infer the world-line homology from knowledge of the particle configuration at discrete times. 

For some models, such as color codes \cite{BM07a}, there is no known efficient algorithm that can infer the particles' world-line homology. Other models, such as Kitaev's code, admit an efficient algorithm---the {\em minimum-weight perfect matching algorithm} of Edmonds \cite{E65a} (PMA)---that solves this problem in time proportional to $\ell^6$ where $\ell$ is the linear size of the system \cite{DKLP02a,H04a}. With such a scaling, it is not possible to handle lattices of more than a few hundred sites long in any reasonable time, ruling out any practical application in this context. Additionally, this algorithm is sub-optimal. As we show below, decoding a topological code reduces to minimizing the free-energy over all homology classes of the system. The PMA minimizes the energy instead, which can lead to additional corruption of the information.

In this Letter, we present a real-space {\em renormalization group} (RG) algorithm to accomplish this task in time $\log \ell$ (based on a parallel architecture, or $\ell^2\log \ell$ in series). The algorithm is not exact and makes use of mean-field equations implemented as a  {\em belief propagation} algorithm (see \cite{Yed01a}). Despite these approximations, our algorithm achieves a successful decoding of Kitaev's model up to a tolerable noise threshold (diffusion rate) that exceeds the one obtained by the PMA. Our algorithm is suitable for a large class of topologically ordered systems including color codes and can be adapted to surfaces of arbitrary geometries and topologies.

Our algorithm is also of use for fault-tolerant quantum information processing schemes based on topological error correcting codes \cite{DKLP02a,RH07a,BM09a,FSG08a}. This method achieves a high (0.75\%) fault-tolerant quantum computing error threshold by making use of some of the ideas outlined above to encode and manipulate the quantum information in a planar architecture, but do not require a topologically ordered phase of matter. In this setting, our algorithm becomes an efficient decoder, inferring the most likely logical recovery given the error syndrome.  

Finally, our algorithm should be of interest in condensed matter more generally because it provides a numerical method to study topological to quenched disorder phase transitions. We note that real-space RG methods have been devised for systems with topological order \cite{AV08a}, but to our knowledge they do not apply in the presence of disorder. 
\medskip

\noindent\textit{Toric Code}---In the following, we focus on Kitaev's toric code. It consists of a $\ell\times\ell$ two-dimensional square lattice embedded on a torus. Qubits are located on the edges of the lattice.  For each site $s$, define an operator $A_s$ and for each plaquette $p$ (site of the dual lattice), define an operator $B_p$ by
\BE
	A_s=\bigotimes_{q\in n(s)}{X}_q, \quad B_p=\bigotimes_{q\in n(p)}{Z}_q
\EE
where $n(s)$ consists of the 4 neighboring qubits of site $s$, $n(p)$ of the 4 neighboring qubits of plaquette $p$ and $X$ and $Z$ are Pauli matrices. We refer to these operators collectively as {\em stabilizer generators}.

The toric codespace is the ground state manifold of the Hamiltonian: 
\BE
	\mathcal{H}=-\sum_s A_s-\sum_p B_p
\EE
Observe that the stabilizer generators all commute with one another. Thus, the ground space of $\mathcal H$ is their simultaneous +1 eigenspace. This implies that the stabilizer group, formed of arbitrary products of its generators $\{A_s,B_p\}$,  acts trivially on the codespace
\BE
O\ket{\psi_{\mathrm{GS}}}=\ket{\psi_{\mathrm{GS}}} \quad O\in \langle\{A_s,B_p\}\rangle.
\EE

Because they consist of unit loops, one can view the $B_p$ ($A_s$) as generators of the group of  contractible loops on the (dual) lattice. Thus, any contractible loop of $Z$ ($X$) operators on the (dual) lattice acts trivially on the codespace. In fact, any operator that maps the codespace to itself must consist of closed loops of $X$ (or $Z$) operators because, in order to commute with all $A_s$ ($B_p$), they must intersect every site (plaquette) an even number of times. We conclude that only non-contractible loops can map the ground space to itself in a non-trivial  manner.  More specifically, for two non-contractible loops on the lattice, $L_1$ winding around the hole and $L_2$ winding around the body of the torus, and two non-contractible loops on the dual lattice, $L_1'$  winding around the body and $L_2'$  winding around the hole, these operators are 
\BE
\overline Z_{i} = \prod_{q\in L_i}Z_q, \quad \overline X_{i'} = \prod_{q\in L_i'}X_q.
\EE
One can easily verify that these {\em logical operators} commute with the stabilizer group and obey the canonical commutation relations $[\overline X_i,\overline X_j] = [\overline Z_i,\overline Z_j] = 0$ and $\overline X_i\overline Z_j = (-1)^{\delta_{ij}} \overline Z_j\overline X_i$. Thus, these four loop operators form a Pauli algebra of 2 effective qubits \textit{encoded in the topological degrees of freedom} of the ground space. The different combinations of these operators generate the 16 different homology classes that need to be distinguished in order to prevent the corruption of the information.
\medskip

\noindent\textit{Errors \& Decoding}---To characterize the error-correcting capacities of this code, we model the noise by the depolarizing channel. This is a noise model where each qubit independently gets randomized with some probability $q$:
\BE
\rho \rightarrow (1-q) \rho + q \frac{\II}{2} = (1-p)\rho + \frac p3 (X\rho X + Y\rho Y + Z\rho Z)
\label{eq:depol}
\EE
where $p = \frac{3q}4$. As seen from the r.h.s. of Eq.~\eqref{eq:depol}, we can equivalently describe this noise model as acting trivially on the qubit with probability $1-p$, or otherwise randomly applying one of the three Pauli operators. 

Because it anti-commutes with its two neighboring plaquette operators $B_p$, an $X$ error will add two units of energy to the system. We say that it causes two plaquette defects, which can be interpreted as particles. Similarly, a $Z$ error creates a pair of site defects and a $Y$ error creates a pair of both defects. In the case of multiple errors, the defects appear only at the endpoints of error chains. Thus, distinct error chains with the same endpoints cannot be distinguished given the defect pattern they generate. If the union of two such error chains forms a contractible loop, their action on the codespace is identical, i.e. their product belongs to the group of contractible loops. Such errors are said to be {\em degenerate}. If, on the other hand, the union of two such error chains forms a non-contractible loop, they have a distinct effect on the codespace. Thus, the decoding algorithm must somehow decide the most likely homology class of the error based on the knowledge of the defect pattern, also called the {\em error syndrome}.

We now show how this inference problem can be mapped to a statistical mechanics problem \cite{DKLP02a}. Associate to each error chain $c$ an energy $E(c) = J|c| + J'n_{\rm{defect}}(c)$. Here, $|c|$ denotes the weight of the error chain, i.e. the number of physical qubits on which it acts non-trivially. The second term $n_{\rm{defect}}(c)$ counts the number of mismatches between defects and endpoints of $c$. The ratio $J'/J$ encodes the level of confidence of our defect detection.  Then, given the depolarizing error model, the probability of an error chain $c$ given a fixed defect pattern is $Pr(c) = \frac{e^{-\beta E(c)}}{\cZ(\beta)}$ where the partition function is $\cZ(\beta) = \sum_c e^{-\beta E(c)}$ and the Nishimori inverse temperature is given by
\BE
	\beta=\frac 1J\ln\frac{3(1-p)}{p}.
\EE
In the following, we take $J' \rightarrow \infty$, meaning that we only consider error chains $c$ in perfect agreement with the observed defect pattern.

The decoder's task then consists in evaluating the probability of each of the 16 homology classes $\Omega$, generated from all combinations of the 4 non-contractible loops:
\BE
Pr(\Omega) = \sum_{c\in\Omega} Pr(c).
\EE
We can express this probability in terms of thermodynamical quantities. Let $Pr(c|\Omega) = \frac{Pr(c)}{Pr(\Omega)}I_\Omega(c)$ denote the probability of an error string $c$ conditioned on homology class $\Omega$, where $I_\Omega$ is the indicator function of $\Omega$. The entropy and average energy associated to this conditional probability are given by $S(\Omega) = -\sum_{c\in \Omega} Pr(c|\Omega)\ln Pr(c|\Omega)$ and $E(\Omega) = \sum_{c\in \Omega} Pr(c|\Omega)E(c)$ respectively. In terms of the Gibbs free energy $\beta F(\Omega) = \beta E(\Omega) - S(\Omega)$, we obtain
\BE
Pr(\Omega) = \frac{e^{-\beta F(\Omega)}}{\cZ(\beta)}.
\EE
Thus, we see that the optimal decoding consists in choosing the homology class $\Omega$  that minimizes the free energy.

\BF
	\includegraphics[scale=0.4]{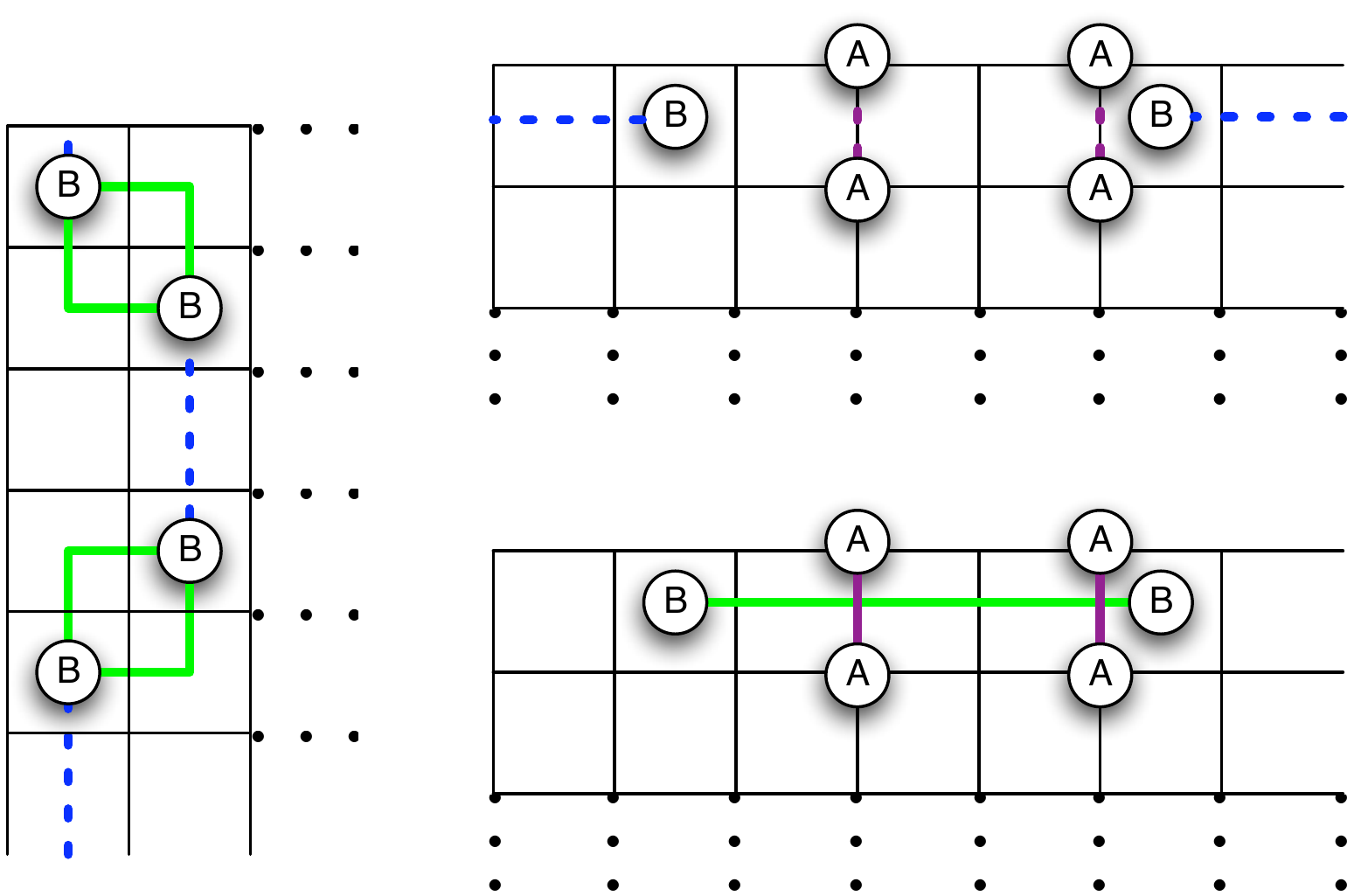}
\EF{Simple defect patterns illustrating the importance of the entropic term or error degeneracy (left) and correlations of errors (right).}{DegCor}

In those terms, we can clarify why the PMA is sub-optimal. By finding the shortest error chain compatible with the error syndrome, PMA minimizes energy rather than free energy. In other words, it operates at zero temperature rather than the Nishimori temperature. Figure \ref{DegCor} (left) illustrates a situation where the entropic term influences the decoding. Namely, two distinct homology classes (dash and full lines) have the same minimum energy configuration, but one of them has higher entropy because it contains 4 degenerate errors, and therefore should be chosen for the decoding. In that case, the PMA has a smaller success probability than a free energy minimization. Repeating this error pattern leads to situations where the difference in success probability can be arbitrarily large. 

Another limitation of the PMA is that it must consider $X$ and $Z$ errors independently. However, because a $Y$ error is the product of a $X$ and a $Z$ error, the two types of errors are highly correlated in a depolarizing model.  Figure \ref{DegCor} (right) illustrates a situation where these correlations influence the decoding. Specifically, the top and bottom figure illustrate two possible error chains for a defect pattern. If the $X$ and $Z$ errors are treated as independent, then the upper configuration has lower energy. Taking into account the correlations of the depolarizing channel on the other hand favors the lower configuration. Again, more elaborate examples lead to situations with highly different success probabilities of the two methods.
\medskip

\noindent\textit{New Approach}---The main idea behind the approach we propose is to approximate the toric code by a concatenated code. A concatenated code is constructed by encoding, say, one qubit into  a code block of $n$ qubits ($n$ should be relatively small), and then encode each of these qubits into $n$ qubits, etc. Clearly, the final number of qubits used is exponential in the number of concatenations, but the failure probability decreases doubly-exponentially. 

Concatenated codes can be efficiently and optimally decoded using a recursive algorithm \cite{Pou06b}. First, we compute the probability of each logical operator (or homology class for topological codes) of the codes appearing in the last concatenation step. As in the previous section, this problem maps to the evaluation of a free energy for a system of finite size $n$, so can be solved by brute force. These probabilities become the noise model for the second-to-last layer of codes, replacing the depolarizing channel. In turn, these code blocks can be decoded by brute force. This process is repeated until we reach the top layer of the code, providing a probability distribution for the net error affecting encoded qubits. 

One can view each step of this decoding process as a RG transformation. The noise model evolves from one concatenation level to the next based on some simple procedure. The transformation is not homogenous in space and time, reflecting the inhomogeneity of the observed error syndromes. 

It is noteworthy that all the codes appearing in a given concatenation layer can be decoded simultaneously. Hence, this decoding scheme is tailored for parallelization and therefore has a running time proportional to the number of concatenation layers, i.e., scaling logarithmically with the number of qubits. The time to decode a single block scales exponentially with $n$ (small). 

The toric code is not truly a concatenated code but can be viewed as one if we allow the different code blocks of a given concatenation layer to share qubits, i.e., to overlap with one another. One possible way of breaking the toric code into overlapping code blocks is illustrated on the l.h.s. of Fig.~\ref{SubCode}. The dashed edges represent qubits that are shared by two neighboring code blocks. 

Each of these code blocks is a small open-boundary topological code, so it can be decoded by brute force. The outcome is a probability distribution for the two logical qubits encoded in this small code. This RG procedure, schematically illustrated at Fig.~\ref{SubCode}, is the core idea behind our method. It maps an error model on 8 qubits to a renormalized error model on 2 qubits. The procedure is repeated by joining 4 pairs of such renormalized qubits to obtain a new block of 8 qubits. At the final iteration, we obtain the error distribution of the two qubits encoded in the toric code. 

\BF
	\includegraphics[scale=0.35]{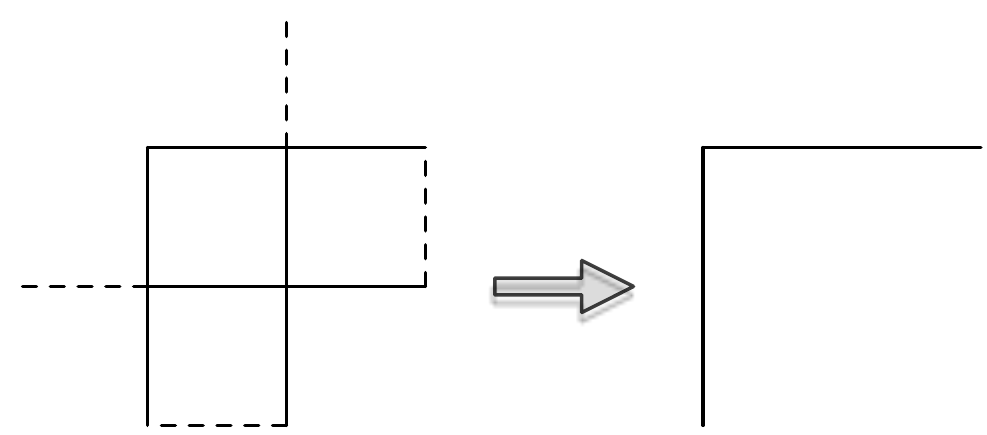}
\EF{The 12 qubits that constitute a code block in the concatenation approximation. They form a surface code. Dashed links represent qubits shared between blocks.}{SubCode}

\noindent\textit{Results \& Improvements}---We use Monte Carlo sampling to characterize the decoder's performance. The basic RG decoder described above achieves a depolarizing error threshold $p\sim7.8\%$ compared to $p\sim15.5\%$ obtained using the PMA \cite{HPP01a,H04a}. This threshold can be significantly improved  by imposing self-consistent conditions on the qubits that are shared between blocks. Suppose qubit $Q$ is shared between a block on the left and a block on the right. Let $D_L$ and $D_R$ denote the defect pattern on the left and right code block respectively. Each block will assign a conditional error probability to $Q$ $Pr(E_Q|D_R)$ and $Pr(E_Q|D_L)$ that are different in general. To improve our decoding algorithm, we impose that all such conditional probabilities agree. This leads to a set of mean-field self consistency equations that we solve by a belief propagation algorithm. We obtain a threshold $p\sim 15.2\%$, quite comparable to the one of the PMA. 
 
\BF
	\includegraphics[scale=0.55]{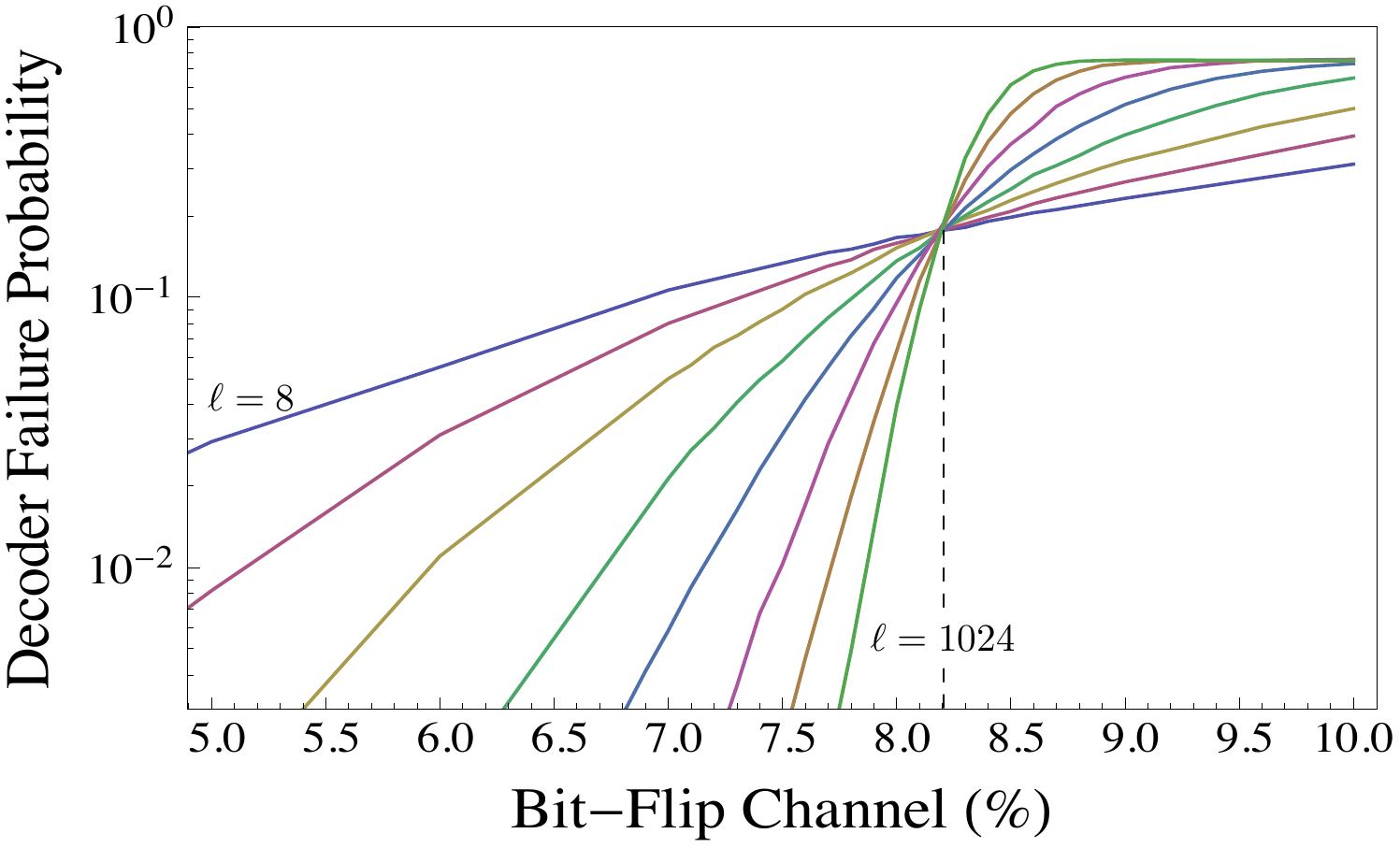}
\EF{Failure probability as a function of bit-flip probability $p$ for a $\ell=8,16,32,\ldots 1024$ toric code decoded using the $2\times 1$ block RG algorithm with 3 belief propagation rounds. Each data point is $10^4$ trials.}{BP}

Our algorithm is very flexible and allows for  multiple tradeoffs between noise suppression and running time. Note that the running time of all the variations we present scales like $\log \ell$, but the constant pre-factors greatly differ. For instance, we can replace the $2\times 2$ code block of the RG scheme by a $2\times 1$ code block. This slightly deteriorates the threshold to $p\sim 12.9\%$, but speeds up the decoding by a factor of a few hundred. We can also choose to ignore the correlations between $X$ and $Z$ errors and decode them separately as bit-flip and phase-flip channels.  This is a more direct comparison with the PMA. Our method yields a threshold of $p\sim 9\%$ with $2\times 2$ RG blocks and $p\sim 8.2\%$ with $2\times 1$ RG blocks (see Fig.~\ref{BP}), compared to $p\sim 10.3\%$ for the PMA, and speeds up our RG algorithm by a factor of a few thousands. Even without parallelization, this enables us to decode lattices a thousand sites long, c.f. Fig.~\ref{BP}.

\BF
	\includegraphics[scale=0.55]{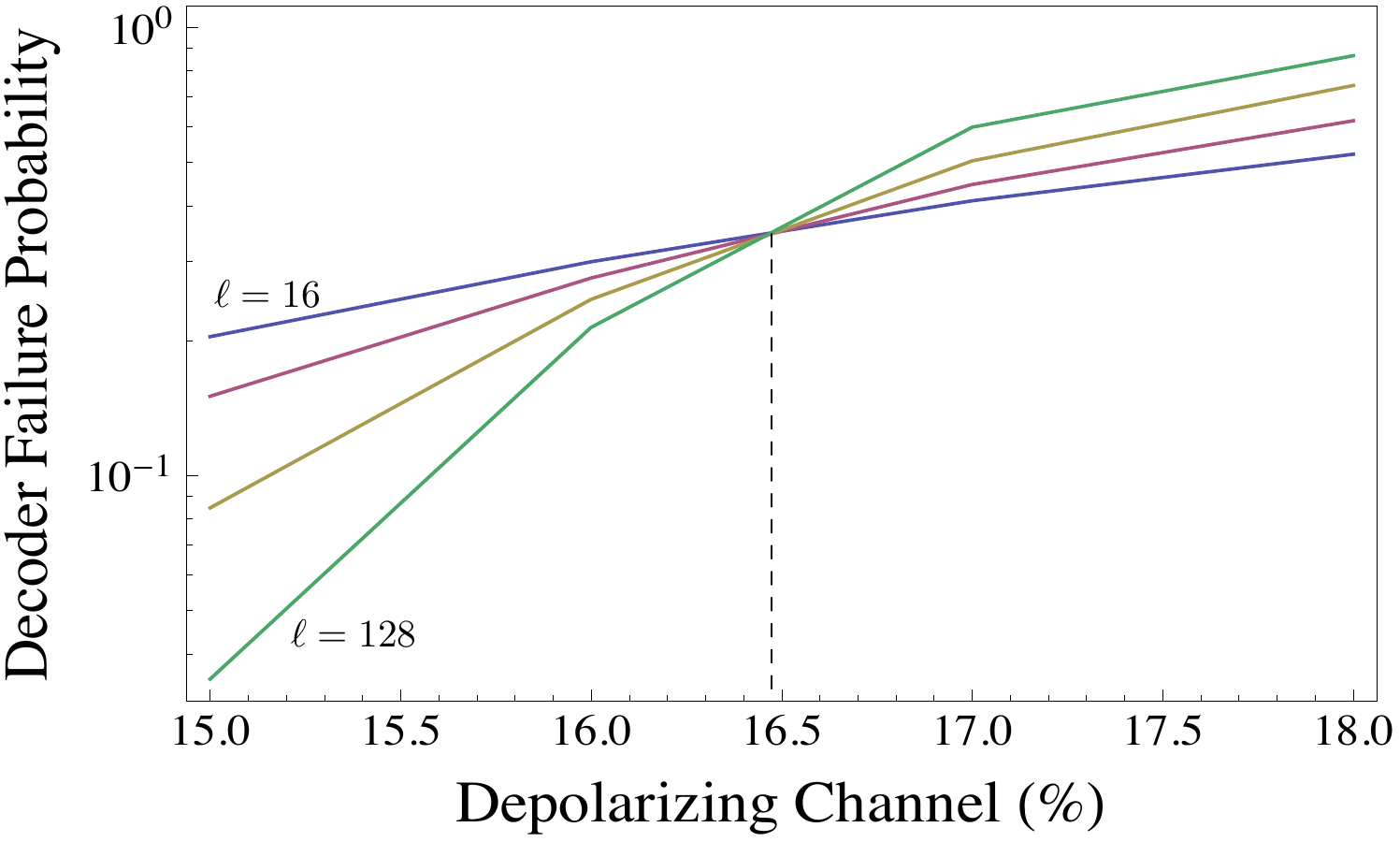}
\EF{Failure probability as a function of depolarizing strength $p$ for a $\ell=16,32, 64$, and $128$ toric code decoded using the $2\times 2$ subcode RG algorithm with three rounds of belief propagation with prior physical decoding.}{ResCurves}

We can further improve these thresholds by combining the RG scheme with the sparse code decoder we presented in \cite{PC08a}. We first use the sparse code decoder with the full (correlated) error model to update the $X$ and $Z$ error probabilities on each physical qubit. We then use the RG scheme to decode $X$ and $Z$ errors independently. The threshold achieved is $p\sim 16.4\%$ (see Fig.~\ref{ResCurves}), exceeding the one of the PMA. 
\medskip 

\noindent\textit{Conclusion}---We presented a real-space RG method to decode the quantum information stored in the topological degrees of freedom of a system in the presence of noise. Our method is very versatile, faster than existing schemes, and tolerates a higher noise rate. It is also the first known algorithm to decode the color code. Beyond its immediate implication to quantum information, it opens the door to new numerical methods to study topologically ordered systems with quenched disorder (see \cite{BP09a} for a similar combination of RG and belief propagation in the context of condensed matter physics).

\noindent\textit{Acknowledgements}---We thank Jim Harrington for many stimulating discussions. This work was supported in part by MITACS, NSERC, and FQRNT. Computational resources were provided by the RQCHP.



\end{document}